\documentclass{llncs}

\usepackage{smwifclass}
\usepackage[utf8]{inputenc}
\usepackage{amsmath}
\usepackage{amsfonts}
\unlessclass{llncs}{\usepackage{amsthm}}
\usepackage{algorithm}
\usepackage[endLComment=~]{algpseudocodex}

\usepackage{balance}
\usepackage{xcolor}
\usepackage{url}
\usepackage{smwmath}
\usepackage{smwtools}
\showdetailstrue
\showanotatrue
\begin{IfClassAcm}
  \setcopyright{acmcopyright}
  \copyrightyear{2023}
  \acmYear{2023}
  \acmDOI{XXXXXXX.XXXXXXX}
  \acmConference[ISSAC '23]{Accepted to ISSAC 2023}{XXX NN-NN, 2023}{Troms\o, Norway}
  \acmPrice{15.00}
  \acmISBN{XXX-X-XXXX-XXXX-X/XX/XX}
\end{IfClassAcm}
\begin{document}
\title{\mbox{A Symbolic Computing Perspective} \mbox{on Software Systems}}
\author{
  Aurthur C. Norman$^1$ 
  \and 
  Stephen M. Watt$^2$
}
\institute{
  Trinity College,
  Cambridge CB2 1TQ, UK\\
  \email{acn1@cam.ed.uk}\\
  ~
  \and
  Cheriton School of Computer Science,
  University of Waterloo, N2L 3G1 Canada\\
  \url{https://cs.uwaterloo.ca/~smwatt} \\
  \email{smwatt@uwaterloo.ca}
}
\maketitle
\begin{abstract}
Symbolic mathematical computing systems have served as a canary in the coal
mine of software systems for more than sixty years.  They have introduced
or have been early adopters of programming language ideas such ideas as
dynamic memory management, arbitrary precision arithmetic and dependent
types. These systems have the feature of being highly complex while at the
same time operating in a domain where results are  well-defined and clearly
verifiable. These software systems span multiple layers of abstraction with
concerns ranging from instruction scheduling and cache pressure up to
algorithmic complexity of constructions in algebraic geometry.  All of the
major symbolic mathematical computing systems include low-level code for
arithmetic, memory management and other primitives, a compiler or
interpreter for a bespoke programming language, a library of high level
mathematical algorithms, and some form of user interface.  Each of these
parts invokes multiple deep issues.

We present some lessons learned from this environment and free flowing
opinions on topics including:\\[0.5\baselineskip]
{
\raggedright
\begin{itemize}
\item  Portability of software across architectures and decades;

\item  Infrastructure to embrace and infrastructure to avoid;

\item  Choosing base abstractions upon which to build;

\item  How to get the most out of a small code base;

\item  How developments in compilers both to optimise and to validate code have always been and remain of critical importance, with plenty of remaining
challenges;

\item  The way in which individuals including in
particular Alan Mycroft who has been able to span from hand-crafting
Z80 machine code up to the most abstruse high level code analysis
techniques are needed, and

\item  Why it is important to teach full-stack thinking to the next generation.
\end{itemize}
}

\end{abstract}
\thispagestyle{plain}\pagestyle{plain}

\section{Introduction}
This paper is to celebrate and illustrate some of the threads that have
pervaded Alan Mycroft's work across the years. He has combined skill in
the sort of low level detailed code constructions covered in Hakmem~\cite{hakmem} and the Hacker's Delight~\cite{HackersDelight-1stEd,HackersDelight-2ndEd} with some of
his skills there honed through his support for the BCPL compiler for the Z80
and also on perhaps particularly the string-processing parts of the C
library that accompanied the Norcroft \texttt{C} compiler~\cite{Norcroft}. But he
has also put much energy into higher level code optimization using
various styles of global analysis, so it is proper to look at the
consequences that flow from developments in that area. He has also
focused on ways that compile-time analysis can help with getting code
correct - some of that will have grown out of association with Edinburgh
Standard ML~\cite{EdinburghML} where one starts with a language where
naive implementation may be inefficient but where the semantics are clean
enough that even radical code transformations can be safe. A ideal that
flows from that language is that users should express their needs with
the very greatest clarity and generality and that the language
implementation should (ideally!) arrange that performance ends up
satisfactory.

We try here to look at an area that illustrates the fact that if one
looks at a problem from only a single perspective, only thinking in terms
of a single level of abstraction, then important aspects of a practical
implementation will be lost. And a part of what we observe is that
developments in optimizing compilers over the last say 30 years have made
it possible to express code in a much cleaner way that had previously
been possible. Our case study is the apparently simple task of long
division of multi-precision integers.

In many contexts algorithmic performance is looked at through the prism
of big-O notation. It is only in special areas the improvements by less
than perhaps a factor of 2 are viewed as worthy of discussion, and
speeding things up by just a few percent is usually though of as
immaterial. But Alan has had involvement in precisely such special areas:
one is compiler optimisation where over the years a succession of
incremental improvements make life better for everybody. The second is in
low level libraries where even small enhancements (for instance in being
able to copy strings using word not byte operations) helps all users.
Finally, as compilers improve, those limited areas where machine-code
implementation wins have shrunk, and compiler-based analysis has allowed
use of higher levels of abstraction in even performance critical code.
This higher abstraction has improved the prospect of static reporting of
errors as well. Alan has contributed either directly or indirectly to all
of these.

The particular case study we cover here is at almost the lowest level of
any symbolic computation systemķ- arithmetic. Software that is to
manipulate algebraic expressions needs arbitrary precision integer
arithmetic. Unlike the situation for those concerned with cryptography
this has to cover numbers whose size can not be predicted in advance.
Unlike the situation for those concerned with breaking records for high
precision values of elementary constants it almost never calls for the
ultimate asymptotically best algorithms. Because this arithmetic
underpins all other activity it even small improvements to it impact the
whole system.

The current market-leader for big integer arithmetic is ``GMP'' -- the
GNU multi-precision library~\cite{gmp}. It is notable as a widely
used library that still has heavy
reliance on machine code to achieve its objectives. The first version
of GMP was released only in 1991, well after we became involved in
implementing arithmetic, and so we started from code-bases and mind sets that
predated it. Since then we have valued abstraction and readily
portable code (GMP has separate bodies of assembly code for perhaps a
couple of dozen architectures, all calling for maintenance) above
absolute performance. But recently we have wanted to consider two
questions
\begin{enumerate}
    \item How close to GMP performance can high level code now come?
    \item Are the algorithms from Knuth volume 2 still close enough
          to the the last word?
\end{enumerate}
Our stance is also that low-level underpinnings for a system deserve
serious review every decade or so, this view having emerged over
around 50 years of experience.

\section{Acceptable Absolute Performance Given Good Compilers}
Going back to the times when Alan Mycroft was a student, one of us (ACN)
had multiple-precision arithmetic using base $10^9$
(arguing that a power of 10 base let input and output happen in linear
time but left the asymptotic cost of arithmetic unchanged).
Around 1989 that was replaces by a version using $2^31$ and coded in \texttt{C}
rather that BCPL.
Twenty five or so years later that needed to be replaced by a
version using 64-bit digits and built using \texttt{C++}.
These were not capricious changes -- one was triggered by the need to move
from a BCPL base to a \texttt{C} base
(for instance using Norcroft \texttt{C}~\cite{Norcroft} developed along with Alan!)
and the other to keep up to date using \texttt{C++} in a 64-bit world.
Many serious long-lasting systems will need their foundations
dramatically re-worked over similar time-scales.

Meanwhile the other of us (SMW) became involved in the
Scratchpad~\cite{scratchpad}, Axiom~\cite{axiom} and Aldor~\cite{asharp,aldor} work at IBM.
Coincidentally at its very start Scratchpad included serious parts of
Reduce~\cite{reduce} which was the driver for the first work thread.
Axiom was seriously concerned with generality, and this led SMW to
consider both the extent to which big-number code could be written
neatly parameterised by the width of digits, and how the algorithms
used could be expressed so as to be directly usable in a range of
other domains that were not purely numeric. His sequence of papers on
``shifted inverses''~\cite{watt-issac-2023,watt-casc-2023} triggered the current work.

Pleasingly, by using \texttt{C++} templates to achieve specialization for small cases
and with the highest level of optimization using current compilers -- and
a certain amount of care reading the Hacker's
Delight~\cite{HackersDelight-1stEd} -- the unsigned multiplication code written
in \texttt{C++} could match or even beat GMP multiplying numbers with up to around
130 decimal digits. From there on gpm starts to win mostly by a factor of
up to 2. This suggests that there remains some scope for further compiler
development to help us claw back that discrepancy! At some stage beyond
where Karatsuba~\cite{karatsuba} becomes the preferred scheme GMP switches
to the Toom~\cite{gmp} family of algorithms, before eventually moving to
use of the FFT. Rather than that, the competing \texttt{C++} code switches to using
three threads for the three top-level sub-multiplications that Karatsuba
performs, making that transition once thread management overhead is
properly balanced by concurrency savings. The means that comparison
between the two platforms is not quite straightforward, but of the
measurement is of elapsed time on an otherwise lightly loaded system it
allows the portable \texttt{C++} code to be at least competitive against GMP out
to well over 30000 decimals. This is typically as far as general purpose
symbolic computation cares for -- calculations needing more than that are not
liable to terminate in a sensible amount of time anyway. This confirms
what one might have hoped, that modern compiler optimisation can diminish
the need for assembly code even in extreme cases. But it also confirms
one's uncomfortable suspicion that it is not yet perfect so continued
work in the area is called for.

\section{Long Division}
With multiplication code stable the and prompted by the SMW work on
shifted inverses attention re-focused on division, and this forms the
main core of this paper.

Way back in 1969 Knuth explained to the world how to do fast long
division. His procedure is based on wanting to to compute a quotient
$q = u/v$ of $N$ digit numbers. Since he is actually considering
fractional values rather than integers all the numbers concerned have
just $N$ digits and his explanation will be most directly relevant to
the implementation of high precision floating point.

Jebelean~\cite{Jebelean-div2} in a paper on practical integer division stated
in the introduction to his report that a scheme designed along the lines
of the Knuth method would lead to long division being around 30 times as
costly as a multiplication. This would render it of little practical
value. We set out to see if use of a modern code-base (Jebelean's
arithmetic used a base of $2^{29}$), shifted inverses and a fresh round
of consideration could change that judgment.

Knuth starts by computing $w = 1/v$ using an iteration
$w \leftarrow w + w \times (1 - v \times w)$
performing only the last step to full precision $N$. The previous to $N/2$
and so on. The total cost of finding the reciprocal is then $4 \times M(N)$
where $M(N)$ is the cost of a single multiplication of two $N$ digit fractions
to obtain an $N$ digit result.

Next compute $q^\prime = u \times w$ and it is an approximation to the final
result and with care it will be correct within 1 and can only
be an underestimate.  Finally compute $r=u-v\times q^\prime$ and compare 
against $v$ -- if necessary do a minor correction. In all this has
used time $6 \times M(N)$ [plus some linear cost work]. Students have been
taught this scheme for generations. but in reality every single
simple tiny step has greater depth.

We will look at this in detail and show the extra considerations that emerge
when looking at the algorithm from a perspective where absolute rather than 
just asymptotic costs matter. We also have some comments on code
expressiveness and compile-time validation and to make, and believe that
both concern for fine-detail issues that impact performance and higher
level ones relevant to correctness and maintainability are relevant for
this meeting. 

A first thing to note is that Knuth's explanation is in terms of
numbers with an implicit leading binary point, in other words fractions.
Even though at times this may still provide a convenient way to think,
it means that the code for integer division diverts into a different
domain for much of what is done. SMW properly viewed this as unsatisfactory
not just on aesthetic grounds but because it complicates any attempt to have
one body of code (with an associated single proof of correctness) that is
applicable across multiple domains. Axiom and Aldor try hard to keep
all code such that the characteristics of underlying domains (rings,
fields, vector spaces, polynomials, non-commutative versions of all
those\ldots) can be use to parameterise algorithmic code. He set about
re-formulating fast division so that all intermediate values where in the
same domain as the inputs, but ``shifted''. The work reported here started
with the idea of implementing fast division within the ACN code body
(which just used classical methods for that operation) both to enhance that
code and to further test and demonstrate the SMW variant of the algorithm.

But then we next note that given that calculation is going to be done 
using digits (in this case 64-bits wide) rather than just abstract numbers
it is desirable to align all values to make full use of each digit. Thus
the notation $1/v$ is taken as an invitation to left-shift $v$ until its
top digit is almost full. Informally this can be thought of as 
normalization to the range [0.5, 1). Then we compute 0.5 divided by this
shifted value rather than $1/v$ and obtain a nicely normalised reciprocal
also in the range [0.5, 1] - save for the case where $v$ is exactly a power 
of 2 which just gets treated specially. Of course the shifting has to be
allowed for am in effect undone at a later stage, but that is easy so I 
will not mention it again, even though it represents a number of extra
lines of code. Also the explanation here in terms of fractional values
has to be interpreted as talking about integers with associated shift values.
This in not quite like use of floating point where every intermediate value
tends to be re-normalised to have its top bit set, and it is not
quite like most scaled arithmetic were there will often be scaling by
a fairly fixed amount -- here as we go the amount of shifting will end up
varying from step to step.

In the calculation $q = u/v$ many in the past have spoken of
working with $N$ digits. Well $u$ and $v$ in general have unrelated sizes, 
so a single parameter $N$ here is an over-simplification. It is more 
probable that the sized of $v$ and $q$ are what should be thought about
and performance predictions can not be uni-dimensional.

In real code if numbers are reasonably small it will be proper to drop back to
classical methods. When an iteration is called for a starting
approximation will be needed and there has to be a judgment about how
accurate that should be because even if a single digit would suffice
it is not obvious that such a choice will be best.

So we now consider the important iterative step
$w \leftarrow w + w \times (1 - v\times w)$ and we assert that it has
issues in every sub-operation!

At any stage we want to calculate using only values that are meaningful 
and we only want to generate outputs that will be necessary. So consider 
the inner $v \times w$. At any stage w will have (say) $k$ digits correct,
so there is no merit in looking at more digits of $w$ than that.
However the updated value we are computing will have $2k$ digits correct,
and those depend on $2k$ 
digits from $v$. So in that multiplication we multiply a $2k$ digits value by 
a $k$ digit one and that yields a $3k$ digit result. However because $w$ is 
already a $k$-digit approximation to the reciprocal of $v$ we know in advance 
that (almost) the top $k$ of those digits will exactly cancel with the 1, 
and so those do not need computing. Furthermore (almost) the lower $k$ 
digits of that result are not needed because they would contribute to 
digits beyond $2k$ in the next value of $w$. So what we need is an unusual 
form of multiplication that forms the product of a $2N$ digit number by an $N$ 
digit one and just return the middle $N$ digits on the result. But it is 
even messier because we actually need those middle $N$ digits with extra 
guard digits surrounding them!

These cases of looking at just some of the digits of a product and the
issues of looking at different numbers of digits within different values
amount to just what Watt was expressing at a higher level on his papers. They
are explained in grim detail here to illustrate just how much care has to
be taken throughout the implementation, and how much might be lost by
viewing everything at a higher and more mathematical level.

Now consider the ``$w + w\times$'' part. The multiplication is of
a $k$ digit $w$ by only $k$ 
digits from the term just computed because it is (almost) the case that 
this is not added to the existing value if $w$ but merely concatenated on its 
end to (almost) double the number of correct digits. So only (almost) 
the top $k$ digits of that product are required. There are two driving 
issues behind the repeated uses of the word ``almost'' here. One is that when 
one computes the top half (say) of the product of a pair of numbers by a 
scheme other than forming a full product and discarding low digits there 
are liable to be carries from the omitted low part of the calculation that 
are missed so the value formed will be slightly low. It is therefore 
necessary to bound that level of inaccuracy and maintain guard bits 
sufficient that it does not hurt the end result. The second issue is that 
in the Newton Raphson iteration even if at one step a value may have all 
bits of $k$ digits correct and then at the next we expect to get $2k$ digits 
correct, there can be rounding or truncation errors relative to the full 
result both in the initial $k$ digits and in the calculation that obtains 
the updated value. Tiny errors can escalate. This issue interacts with the 
fact that the eventual number of digits required may not be a power of 2, 
and so for instance if $2k+1$ digits are required in an end result the
iteration that leads to it is liable to be starting with values stored
as $k+1$ digits but where the least significant of those digits could afford
to have almost half its low digits incorrect. This adds extra depth to the 
simple sounding statements about use of precision $N, N/2, N/4\ldots$

$q^\prime=u \times w$ is another places where only high digits from a
product are needed. Well the magnitude of the quotient can be estimated
from the sizes of divisor and dividend. At this stage we only need to use
high digits from $u$, and at this stage we can see that the reciprocal $w$
needed to be computed to a number of digits to match $q$. Note that this
level of precision may be either greater or less than the size of the divisor,
and coping with that adds further detail to the iterative process -- in 
particular the operation described above as multiplying a $2k$ digit by a $k$ 
digit one will sometimes have to allow for (and take advantage of) the 
divisor $v$ not having fully $2k$ digits.

The final step that Knuth presents forms $r = u-v*q^\prime$.
He has arranged that $q^\prime$ is either the correct quotient or just
one too small, and as a result $r$ will be in the range $0 \le r < 2v$.
That means that we can tell in advance 
that many high digits in the subtraction there will cancel exactly, and so 
this is a case where only low parts of the product are required. If we
require the remainder as part of our output this is as far as we can go,
but there are occasions when division is performed and only the quotient 
is wanted. In such cases computing even the full low part of
$w \times q^\prime$ is usually unnecessary.

If we did the full calculation of $r$ and then compared it 
against $v$ that would be done by inspecting its top digit first and only 
working down to check lower digits if the issue had not been resolved. One 
might hope that using 64-bit digits that testing only the top digit would 
almost always be sufficient. That means we may be able to get away with
finding just a single digit from some well-chosen place towards the middle 
of $v \times q^\prime$. This can be done with controllable error in
linear time rather than $M(N)$ time. However three issues intervene.
The first is that the top 
relevant full digit of $w \times q^\prime$ may have either almost all its
bits in use or only a few. In the latter case testing just those few bits does not 
provide as reliable a test as would be ideal. This can be coped with by 
using code that extracts a digit-sized value from a product but aligned
by a bit-address rather than a digit address. This is not obviously a
primitive operation widely discussed in thge literature.
The second issue is that a one-digit part product can have inaccuracies
because carries that would 
have contributed to a perfect value have been missed out. So those errors 
need bounding and allowing for. And finally there is the issue of what the 
nost challenging cases might be. Here that will be when $u-v \times q^\prime$
is very close to $v$ in value, in which case the comparison may not be
resolved by inspecting just one high digit. In particular this can be the
case when the division is going to prove to be exact (and $q^\prime$ was
one too low). Exact division seems a really bad case to have badly
handled in a version of division code that will not be returning a remainder!
To mitigate that we round $q^\prime$ rather than truncate. This means that $r$
has to be tested not just to see if it is less than $v$ but also to see if it
was negative, but it moves the case where this one-digit check for
correction is insufficient from near exact divisions to one where the 
remainder is around $v/2$. Obviously if the one-digit test is inconclusive
it makes sense to drop back and use the original low-half-of-product 
scheme.

All the above makes use of version of multiplication code that deliver 
some but not all the digits of the full result. Mulders~\cite{mulders}
pioneered this. He was concerned with multiplying power series and so naturally he
looked at keeping the $N$ low terms from the product of two series each of 
which had $N$ terms. He showed that he could form the product in time that 
was say 70\% to 80\% of the cost of performimg a full Karatsuba style
multiplication. Hanrot and Zimmerman~\cite{HanZim} considered his scheme
in some depth and in particular looked into the optimal value for his
parameter $\beta$. There were also associated with its use in the \verb+mpfr+
multi-precision floating point library, where they will have just been
concentrating on the top half of a multiplication of two equal-sized
numbers, but they will have considered carry propagation carefully. 

For use here it has been necessary to adapt things so that 
the two inputs do not have to have the same number of digits, to allow for 
carry operations and to have versions that keep flexible numbers of high, 
middle or low digits from the product. That of course all depends on 
having underlying fast full multiplication, and so the Karatsuba 
procedures have to be bolstered with code that allows for inputs not 
balanced in terms of their digit counts. At yet lower levels performance 
can depend on just how carry detection and propagation is implemented, how 
the temporary workspace that Karatsuba and Mulders need is managed and on 
overheads that arise when the resulting library is built so it can work 
in a threaded environment.

As an implementation became close to complete it became possible to start
some performance assessment.

The first observation is that in general the cost of a classical long
division $u/v$ is only modestly greater than that of classical multiplication
of $v$ by the quotient $q$. This should probably not come as a big surprise!
It also makes sense that when the quotient has fairly few
digits any chance for the Newton-Raphson ``fast'' division to shine
has to depend on even $q$ being rather large and hence the two inputs $u$
and $v$ will be enormous.

If the divisor and quotient are about the same size the fast method
can be a winner with the cost of $u/v$ being only about 4 times that of
$v \times q$, but the various overheads mean that with the current code-base
a division of a 100-digit number by a 50-digit one is still faster using
classical methods. Nut the ratio of the cost of clever division to
Karatsuba multiplication is still not much more than 4 -- a long way
short of the 30 that Jebelean had projected, but that of course does not
invalidate the merits of the alternative scheme that he presents.

The case that we had not initially expected arises when the quotient has
many more digits than the divisor and there the Newton-Raphson shows
disastrously poor performance. This is of course because it is needing to
compute a value of $1/v$ to the number of digits precision set by $q$.
It becomes clear that when $q$ is going to be significantly larger than $v$
that the division should be conducted as if by short division by $v$,
partitioning $u$ into appropriate sized super-digits. That will result
in all the internal divisions being of the $2N/N$ variety where the
assymptotically good method actually pays off.

\section{Correctness and Abstraction}

It should be apparent from the above explanation of our division code
that it ends up complicated enough that correctness can not be a given.
In particular are all stages we want to compute with only the minimal
number of digits to maintain accuracy -- the closer to the wind we can
sail the faster the code will be.

There are two components to the task of getting things right. The first
is illustrated by a need for a bound on the error due to ignoring some
potential carries when we compute just the top half of a product.
In some of the work we were involved in properly detailed coding: for
instance we write \verb/(-(from>=M+1))&(from-M)/ rather than
\verb/from>=M+1?0:from-M/ because using the compilers we were the former
tuned into branch-free code and ran measurably faster than the latter. But
then we need to switch into mathematician mode and derive and then
prove a bound on the impact carries could have on a partial product.
We would like to believe that Alan Mycroft is the sort of person with the
breadth to contribute at both end of the span of abstraction levels.

With regard to the low-level hackery we are very aware that compiler
improvements over the years have made it possible to achieve results
that previously called for lower level code. For instance some compilers now
recognize fairly natural-looking idioms and generate machine code that
makes proper use of carry flags and ``add-with-carry'' operations. But
out case where we want to generate branch-free shows that such a line of
work has not fully run its course. But we are also strongly appreciative
of compiler work that in-lines functions, maps variables onto registers
and all the other clever things that allow us to write code in a cleaner
and more abstract way than in the past.

We also note that way on which compilers can increasingly propagate
information through code and and detect issues. We very much want that to
continue to improve. The experience developing this fairly densely
detailed code all intended to implement (in the end) very clear cut
mathematical operations has shown that at least with the compilers
currently in general use there is still plenty to be done. In a better
future world all of the off-by-one and not-quite-enough-bits-for-accuracy
bugs we had to detect by fairly traditional methods might ideally be
spotted based on static code analysis. This of course can involve continuing
the merge of proof technology with compilation.

The final issue of language design and compilation that we feel that this
effort has highlighted for us is the need to be able to express actions at
the highest possible level of abstraction while still maintaining fine
detailed control of issues that impact performance. As an example of a
conflict we faced in this style consider the fact that most of the more
elaborate big-number functions need workspace sized as per their inputs.
A clean way of allocating this space might be use of \texttt{C++ std::vector}, and
then a reasonably plausible compiler can lead to unchecked indexed access
being as efficient as use of simple \texttt{C}-style arrays. However in library code
that is liable to allocate and release memory very frequently the potential
costs of \verb+new+ and \verb+delete+ operations are a worry - leading us
to provide our own scheme exploiting all our understanding of sharing
and lifetime properties of our workspace. We obviously try to implement that
with an interface that makes its use seem as high level and abstract as
possbile (thank you templates, overloading,\ldots) but all in all we find the
gulf between the around 3 lines of explanation that Knuth provides and the
several thousand lines of code we end up rather horrifying. We are very
aware that if we had coded all of this some decades ago it would have been
even worse, but we very much want compiler work (including language
design and code proof technology) to continue even after the retirement of
one of its contributors.

\section{Results}
Part of the thesis behind the paper is that the real world is messier
than the presentations in most research papers. For integer division
we believe that most explanations of procedures have been characterised
by a single parameter that gives ``the number of digits involved''. We
observe that if an $N$ digit number is to be divided by and $M$ digit one
that there are certainly three domains of performance -- the straightforward
one where $N$ is close to $2 \times M$ but also ones where the divisor
or the quotient is much smaller than the dividend. In the extreme cases the
Knuth scheme -- applied in an naive manner -- is unsatisfactory. If the
divisor is small relative to the dividend but large enough to justify
non-classical treatment it is much better to perform an operation in the
style of classical short division treating the divisor as defining a sort
of digit. If the quotient is going to be small a scheme that uses an
iteration to generate a shifted inverse of the divisor is extremely
good provided that all calculation is done only to a precision based
on the number of quotient digits. In this case the dominant cost of the
full calculation will be multiplying quotient by divisor and subtracting to
find the remainder. If the user does not actually need the remainder almost
all of that cost can be avoided most of the time leading to dramatic
savings.

As well as there being thresholds based on the relative magnitudes of divisor
and quotient there also have to be ones that reflect that until numbers become
large there is no merit in abandoning the classical algorithms. Just where
these will lie will depend on the relative performance of the classical
division come used as a baseline and on the fast multiplication used for
larger products. Of course multiplication has performance that varies for
inputs that are not the same size and here waters are further muddied by
the use of Mulders-style multiplication that generates only some of the
digits from a full product. In our case an additional complication arises.
Karatsuba multiplication works by decomposing a product so that to
multiply a pair od $N$ digit values one performs three multiplications
on $N/2$ ones. For large enough $N$ that synchronization overheads are
balanced by concurrency savings these three sub-products are calculated in
separate threads, giving a reduction in elapsed time but not in the total
number of CPU cycles executed. This certainly brings into focus the issue
of whether timing reports should show elapsed or CPU time, and in the latter
case how much system as distinct from user time needs to be accounted for.
It also means that our multiplication cost grows in a somewhat lumpy
way rather than meeting the asymptotic prediction at all early.

So we need per-platform tuning within a Karatsuba multiplier,
for just how Mulders-style decomposition is used for inputs that do
not match in size and when it is not exactly the top of bottom half
of a product needed and for the changeover from classical to notionally
faster division. At this stage we have not completed all that tuning, and
anyway the main focus here is to expose complicated detail rather than to
claim ultimate performance. So we provide here some measurements that can
at least give an idea of behaviour.

Because it our division code site firmly atop multiplication we start with
measurements for regular simple integer multiplication where the two
numbers being combined each have the same number of digits. We report
elapsed time on am Intel i7-8086k system running Windows 10 and using the
Cygwin \texttt{C++} compiler. The bit-patterns of the numbers multiplied are set
up as random in such a way that successive test runs will use different
random seeds (so as to avoid optimization artefects based on the exact
test cases). Timings for our code are reported against those for GMP
and the key inner loop is essentially
\begin{verbatim}
    clk1 = std::chrono::high_resolution_clock::now();
    for (std::size_t m=0; m<REPEATS; m++)
        mpn_mul((mp_ptr)c,
                (mp_srcptr)a, lena*
                    (sizeof(std::uint64_t)/sizeof(mp_limb_t)),
                (mp_srcptr)b, lenb*
                    (sizeof(std::uint64_t)/sizeof(mp_limb_t)),
        for (std::size_t i=0; i<lena+lenb; i++)
            gmp_check = gmp_check*MULT + c[i];
    clk2 = std::chrono::high_resolution_clock::now();
\end{verbatim}
where the value \verb+gmp_check+ both serves to give a weak confirmation
that our results and those from GMP match and to reduce the changes of
an over-enthusiastic compiler omitting everything because its output was
unused. As well as \verb+my_time+ for the multiplication code that
transitions to Karatsuba and to the GMP figure there is a reference
time that is the timing for a simple version of classical long
multiplication with quadratic cost (and no special casing for
short values). Number lengths are expressed in terms of 64-bit digits
so for instance the line for length 1546 relates to multiplying a pair
of integers eacj of around 30000 decimals. The lengths are not all at neat
multiples of ten because this table is extracted from a larger one which
uses a set of samples that grow geometrically not arithmetically.
This is shown in Table~\ref{tbl:MulNN}.
\begin{table}
\caption{Multiplication of two integers of the same length. Times are reported in seconds per multiplication.}
\begin{center}
\begin{tabular}{r@{~~~}r@{~~~}r@{~~~}r@{~~~}r@{~~~}r}
 Length & Our time&GMP time & Ref time & Ours/GMP & Ref/Ours \\
 \hline
      2 &   0.010 &   0.017 &    0.075 &    0.577 &    7.830 \\
      3 &   0.015 &   0.034 &    0.087 &    0.439 &    5.749 \\
      4 &   0.024 &   0.036 &    0.103 &    0.665 &    4.260 \\
      5 &   0.034 &   0.045 &    0.124 &    0.753 &    3.681 \\
      6 &   0.049 &   0.056 &    0.147 &    0.873 &    2.990 \\
      7 &   0.092 &   0.071 &    0.176 &    1.301 &    1.915 \\
      8 &   0.151 &   0.085 &    0.213 &    1.780 &    1.409 \\
      9 &   0.172 &   0.105 &    0.248 &    1.638 &    1.441 \\
     10 &   0.185 &   0.123 &    0.292 &    1.498 &    1.580 \\
     20 &   0.721 &   0.432 &    1.043 &    1.667 &    1.447 \\
     29 &   1.505 &   0.820 &    2.184 &    1.835 &    1.451 \\
     39 &   2.399 &   1.309 &    3.802 &    1.833 &    1.585 \\
     50 &   3.700 &   1.998 &    6.738 &    1.852 &    1.821 \\
     78 &   7.764 &   4.083 &   15.965 &    1.901 &    2.056 \\
    102 &  12.051 &   6.301 &   29.201 &    1.912 &    2.423 \\
    120 &  16.015 &   7.523 &   39.678 &    2.129 &    2.478 \\
    235 &  42.043 &  22.819 &  152.726 &    1.842 &    3.633 \\
    260 &  44.692 &  26.515 &  186.394 &    1.686 &    4.171 \\
    429 &  72.350 &  54.386 &  509.029 &    1.330 &    7.036 \\
    607 &  98.855 &  87.958 & 1024.906 &    1.124 &   10.368 \\
    740 & 120.235 & 120.761 & 1523.879 &    0.996 &   12.674 \\
   1043 & 217.381 & 189.216 & 2939.518 &    1.149 &   13.522 \\
   1546 & 400.686 & 328.835 & 6583.968 &    1.218 &   16.432 \\
\end{tabular}
\end{center}
\label{tbl:MulNN}
\end{table}

We are of course very pleased with the results for length up to 6 (ie around
100 decimals) and for many calculations in Computer Algebra we view that
range as important. We are frustrated that so far we have not been able to
coax our code and compilers into matching GMP speed there. From 20 up we will
be using Karatsuba and at least we manage to be within a factor of 2
of GMP. At around 200 digits (say 4000 decimals) we start to be able to
use concurrency and that keeps us reasonably competitive against GMP
as far as we fuss. Even though by that stage GMP will be using variants on
Toom rather than just Karatsuba. We do mot measure and do not really concern
ourselves cases with millions of decimals.

Next we report on Mulders multiplication, and again to simplify the
presentation we multiply two equal sized random numbers and keep either
the full product or the top half or the bottom half. Our Mulders code is
capable of delivering an arbitrary slice from the product and so any
overhead associated with that generality is present. We use our
Karatsuba-based multiplication as underpinning. We have code that can
produce a slice of digits from a product using simple classical code, so
we compare this, use of Karatsuba to form a complete product (and then
merely discard unwanted digits) and then Mulders. For generating a complete
product our code degenerates to just use of Karatsuba with a few initial
extra tests that are irrelevant by the time Karatsuba makes sense. And
our comparisons show that we can compute the top half of a product in
times broadly similar to those for the bottom half, so the only section
of our full test results included here are for calculating the lower half
of the product of two $N$ digit values.
The results are shown in Table~\ref{tbl:HalfMul}.

\begin{table}
\caption{Time to compute lower half of product of two $N$ place values.}
\begin{center}
\begin{tabular}{r@{~~~}r@{~~~}r@{~~~}r@{~~~}r@{~~~}r@{~~~}r}
 $N$  & Class   &Kara  &   Fast & Kara/Class  & Fast/Class & Fast/Kara \\
  \hline
 10  &   0.07    &   0.10    &   0.07  &   132.57\%  &  96.42\%    & 72.73\% \\
 14  &   0.13    &   0.20    &   0.12  &  148.17\%  &  89.80\%    & 60.61\% \\
 18  &   0.21    &   0.29    &   0.18  &  139.22\%  &  88.17\%    & 63.33\% \\
 20  &   0.25    &   0.36    &   0.22  &  144.48\%  &  88.76\%    & 61.44\% \\
 24  &   0.37    &   0.42    &   0.37  &  114.31\%  &  99.15\%    & 86.74\% \\
 50  &   1.63    &   1.80    &   1.30  &  109.98\%  &  79.73\%    & 72.50\% \\
 55  &   2.04    &   2.17    &   1.59  &  106.62\%  &  77.74\%    & 72.91\% \\
 70  &   3.19    &   3.38    &   2.21  &  106.04\%  &  69.44\%    & 65.49\% \\
 80  &   4.14    &   3.98    &   2.99  &   96.04\%  &  72.09\%    & 75.06\% \\
 90  &   5.29    &   4.87    &   3.72  &   92.02\%  &  70.37\%    & 76.48\% \\
 95  &   5.78    &   4.69    &   4.02  &   81.13\%  &  69.41\%    & 85.56\% \\
100  &   6.39    &   5.65    &   4.44  &   88.53\%  &  69.50\%    & 78.51\% \\
105  &   7.02    &   6.68    &   4.67  &   95.17\%  &  66.47\%    & 69.84\% \\
110  &   7.70    &   6.88    &   5.08  &   89.40\%  &  66.02\%    & 73.85\% \\
135  &  11.51    &   9.71    &   6.65  &   84.30\%  &  57.78\%    & 68.54\% \\
240  &  36.49    &  20.25    &  19.70  &   55.50\%  &  53.99\%    & 97.27\% \\
250  &  39.29    &  21.23    &  20.53  &   54.02\%  &  52.25\%    & 96.72\% \\
300  &  56.69    &  23.96    &  26.36  &   42.27\%  &  46.50\%   & 110.00\% \\
700  & 302.47    &  60.42    &  72.48  &   19.97\%  &  23.96\%   & 119.96\% \\
750  & 359.93    &  63.92    &  81.34  &   17.76\%  &  22.60\%   & 127.26\% \\
1400 & 1228.45   & 177.73    & 180.51  &   14.47\%  &  14.69\%   & 101.57\% \\
1600 & 1562.67   & 261.09    & 207.99  &   16.71\%  &  13.31\%   &  79.66\% \\
1800 & 2033.82   & 337.94    & 314.95  &   16.62\%  &  15.49\%   &  93.20\% \\
10000& 60694.30  &3520.00    &3558.90  &    5.80\%  &   5.86\%   & 101.11\% \\
\end{tabular}
\end{center}
\label{tbl:HalfMul}
\end{table}

It can be seen that for reasonably big numbers the fast methods are
indeed faster than simple classical code, and for huge cases they are better
by a large factor. Up to a couple of hundred (64-bit) digits the Mulders
code delivers a really useful speedup compared against just using
Karatsuba and then throwing away help of the result. But at about the
point where our Karatsuba implementation goes multi-thread that benefit
gets lost -- perhaps because Mulders in its recursion uses a sequence of
smaller full multiplications that do not use parallelism. At present we find
it hard to explain why our implementation of Mulders does not do better
on truly huge numbers. It perhaps means that pragmatically we should
add in additional thresholds so that for numbers with a really large
number of digits it does not try to be clever!
Well out ``top half of product'' code matches the bottom half
performance up to around 70 (64-bit) digits but then degrades in an even
worse way than the bottom-half code, so further tuning is called for --
or perhaps more sophisticated optimization in the compiler.

We collected similar measurements for the ``shifted inverse`` code that
computes a sort of scaled reciprocal. For that we could use our
existing classical long division as a baseline, then code that used
Newton-Raphson but performed all arithmetic to full precision and finally
our version using precision that adjusted as the iteration proceeded and
Mulders multiplication. As an additional assessment for this we timed
multiplying the input by its computed inverse, so we can compare our
best time with that of a single multiplication. Here the table of
results is tidy enough that we do not need to present much of it. Using
restricted precision for intermediate results and Mulders can save serious
amounts of time as against using full precision everywhere (as expected).
The iterative code is never slower than finding the reciprocal using
classical division, but we needed to get as far as 500 digits before
it was faster by a factor of 5. The shifted inverse is found in a time
that in the best cases is slightly under twice as long as multiplication,
is usually between 2 and 3 times and at worst still under 4 multiplies.

By this stage the multiple opportunities to adjust thresholds in the
underlying code made optimization rather hard even on a single system!
Note that in the above comparisions the various simpler schemes were
being used not just as performance baselines but also for confirmation of
output values, and with input data generated with distinct random seeds
for each test run our confidence tends to grow.

Finally we come to division. Here the calculation will be to
evaluate the quotient $Q$ and remainder $R$ when a value $U$ is divided
by $V$. In each case we will use the corresponding lower case letter
for the number of 64-bit digits in a value. We use $u=10000$ as a case
large enough that we can hope that assymptotic effects will dominate,
and then consider $v=100, 500, 5000$ and $9500$. These of course lead
to quotients taking the same range of lengths.

Rather than showing absolute times we report timing ratios. We scale against
the cost of (our code) multiplying $V$ by $Q$, and then report the
cost of repeating that using classical long multiplication, using
classical long division for $U/V$, using our shifted-inverse code to
calculate quotient and remainder and finally our scheme that just
finds the quotient.

Table~\ref{tbl:ScaledMulDiv} shows how behaviour changes fairly radically for divisions where the
numbers have different lengths. The more traditional Table~\ref{tbl:TradTbl} only considers
$2N\times N$ divisions but observes what happens as $N$ varies.

This of course confirms that eventually fast division beats classical, but
the cross over point is perhaps higher than would be nice! The final row here
is the same test as the $5000,5000$ case above and the slightly different
numbers serve to remind us that measurements on modern computers are
subject to all sorts of variation -- an additional obstacle to
refined optimization. In particular cache consequences of running a
single test repeatedly so as to have a long enough time period
to measure risk leading to results that may not be representative
of real-world usage. These figures also make it clear that a version of
division that does not compute a remainder can help with time savings
when working at higher precisions.
\begin{table}
\caption{Scaled multiplication and division}
\begin{center}
\begin{tabular}{r@{~~~~}r@{~~~~}r@{~~~~~}r@{~~~~~}r@{~~~~~}r}
    $v,q$     & 100,9900 & 500,9500  & 5000,5000  & 9500,500  & 9900,100 \\ \hline
classical mul &   1.96   &  7.07     & 24.30      & 6.92      & 1.97 \\
classical div &   2.16   &  6.43     & 20.55      & 5.89      & 2.01 \\
fast div+rem  &   3.99   &  6.86     &  4.16      & 1.32      & 1.04 \\
fast quot only&   4.00   &  6.80     &  3.20      & 0.29      & 0.04 \\
\end{tabular}
\end{center}
\label{tbl:ScaledMulDiv}
\end{table}
\begin{table}
\caption{$2N\times N$ multiplication and division}
\begin{center}
\begin{tabular}{r@{~~~~}r@{~~~~}r@{~~~~~}r@{~~~~~}r}
   size     &     classical &  classical &  fast  &  divide \\
 in digits  &      multiply &    divide  & divide &  no rem \\ \hline
   20       &        1.33   &    3.24    &   5.98 &    5.22 \\
   40       &        1.67   &    2.63    &   4.58 &    3.81 \\
   60       &        1.82   &    2.42    &   3.96 &    3.28 \\
   80       &        2.06   &    2.36    &   4.45 &    3.65 \\
  100       &        2.18   &    2.41    &   4.34 &    3.54 \\
  150       &        2.39   &    2.47    &   4.09 &    3.27 \\
  200       &        2.63   &    2.64    &   4.16 &    3.36 \\
  300       &        4.45   &    4.22    &   6.33 &    5.17 \\
  500       &        7.51   &    6.99    &   7.15 &    5.81 \\
 1000       &       11.39   &    9.79    &   6.05 &    4.96 \\
 5000       &       24.99   &   20.91    &   4.25 &    3.27 \\
\end{tabular}
\end{center}
\label{tbl:TradTbl}
\end{table}

Given that results can be sensitive to the computer used the same tests
were run on a second system. A Raspberry Pi 5 was used with this choice
partly motivated by Alan Mycroft's involvement in the start-up of
Raspberry Pi.

For multiplication on the Raspberry Pi the \texttt{C++}-coded multiplication beats
GPM up to 7 digits, but only by a small fraction and it would probably be
fairer to declare a dead heap. From there on up to around 200 digits the
speed ratio remains at worst 1.4 and our code os mostly no more than
15\% slower than GMP. From 200-1546 digits where we use parallel Karatsuba
beat GMP where the best observation was taking 65\% of the GMP time to
multiply a pair of 740 digit numbers.

The Mulders multiplication to find the low half of a product shows a
more repeatable speed-up so that Mulders costs about 75\% of Karatsuba
for a wide range of number sizes, but again with ugly behaviour where
a full Karatsuba goes parallel but the sub-multiplications done within
Mulders are smaller and do not. In our current implementation Mulders
loses from around 200-3000 digits, but beyond there it starts to be
a winner again - albeit with less stability in the speed ratio. A hypothesis
is that the heavy use of starting and stopping threads makes timings
sensitive to internal timing in operating system thread-scheduling and
given that the subsidiary tasks are fairly small this generates delays.

Shifted inverses show a pattern similar to that observed on the PC.

For division we obtain Table~\ref{tbl:ShiftedDiv}.
Given the significant differences in architecture it is perhaps
amazing how similar to the Intel table this is.
\begin{table}
\caption{Division with shifted inverses}
\begin{center}
\begin{tabular}{r@{~~~~}r@{~~~~}r@{~~~~~}r@{~~~~~}r@{~~~~~}r}
    $v,q$      &  100,9900 & 500,9500 & 5000,5000 & 9500,500 & 9900,100 \\ \hline
classical mul  &  1.93     & 7.57     & 26.71     & 7.38     & 1.69 \\
classical div  &  2.21     & 7.81     & 26.85     & 7.49     & 1.73 \\
fast div+rem   &  4.09     & 7.86     &  4.28     & 1.73     & 0.94 \\
fast quot only &  4.09     & 7.74     &  3.51     & 0.31     & 0.04 \\
\end{tabular}
\end{center}
\label{tbl:ShiftedDiv}
\end{table}

Looking at how well ``fast'' division works as input sizes grow, on the
ARM we have the shifted-inverse based division matching classical
somewhere between 500 and 1000 digits again just as on Intel.

The main observation is that the Raspberry Pi figures as collected on
Linux look a little less scattered and incoherently variable than the
Intel ones on Windows and this may in part reflect the i7-8086k having
a more complicated instruction processing scheme where while
average performance is excellent detailed timing can be very sensitive to
all sorts of hard to predict interactions. 

One thing which is clear from all this is that integer division following
the Knuth explanation can have a cost less than 4 times that of
multiplying back to recover the dividend - at least in the case where
only a quotient is required not both a quotient and remainder. While this
includes the special case of divisions known in advance to be exact it
does not rely on that situation applying. This is very much faster than the
factor of 30 that Jebelean suggested would apply but nevertheless his
paper claims to achieve a factor of about 2 and so would be even better --
future work should implement that and consider how it applies to quotients
other than the tidy $2N\times N$ case. The other thing that emerges is
that our ``fast'' code only really begins to shine for integers sufficiently
large that we would rather avoid them arising in the first place, for
instance by using modular arithmetic with a word-sized modulus. But
despite these practical reservations we are very pleased with the manner
in which this exercise highlights how much depth can arise when one
attempts to optimise even simple-seeming schemes.

\section{Conclusions and Further Thoughts about education}
All the above explain some of the additional mess that a typical
implementer who is keen to achieve high performance may face. A
properly pedantic implementer will be fixated both on performance and on
correctness, so will need to deploy a mathematician's skills and mind-set
to prove that error never intrude as well as a low level hacker's
understanding of performance issues. With caches, multiple-issue CPUs,
speculative execution and various memory models that can impact
how multiple cores may or may not observe that the others are up to this
has continued to become more and more challenging.
Performance engineering can often
involve a desire to sail as close to the wind as possible, and in this
case error bounds on ``top half of an unbalanced product'' will interact
with the exact manner in which errors propagate through the iterative
step, and careful analysis of just how accurate the initial
approximations to the shifted inverse are. So the difficulty of attaining
correctness has perhaps grown too.

Clearly very many programmers respond by taking the line that delivery
of a product on time trumps correctness and correctness trumps performance.
For many purposes their stance is completely proper, but compilers are
libraries represent special cases where correctness is vital and
optimization impacts enough users that it becomes truly important -- perhaps
especially now that machines have by and large ceases speeding up
substantially and their manufacturers chase benchmarks with a combination
of more cores (which will often not all be activated) and with
special instructions to support important but possible niche applications.

We note that, as system developers,
our own approaches start in some sense from opposite perspectives---one as specific as possible and one as generic as possible.  However, we both consider the consequences of decisions on the full software stack, and end up with designs with a great deal of similarity.

Our illustration has involved integer
division, which might be viewed as fairly low level and fundamental
building block where there are carefully documented solutions that
go back at least as far back as
Knuth II~\cite{Knuth2}. But we assert that proper implementation calls
for a style of computer ``renaissance man'' able to span consideration
from mathematical and abstract down to the finest detail of how to
implement carry detection while combining multi-word integers.
And indeed ideally one who could know hoe then to tune code to exploit
the hardly intuitive performance consequences of modern complicated
instruction execution strategies. For the future we can not count on
performance improvements (and hence resource use reduction) based on
raw CPU improvements. We need a new generation of programmers who -- against
the teaching style of several decades -- value compactness and efficiency,
and continued work to keep improving compilers so that they can deliver
that with clearly expressed and source code that can be validated
effectively. If long division is messy consider almost any ``real scale''
challenge! We need more Mycrofts.

\newpage
\IfFileExists{IfExistsUseBBL.bbl}{%

}{%
\bibliography{main}
}

\end{document}